\documentclass[conference,onecolumn]{IEEEtran}

\usepackage{cite}
\usepackage{url}
\usepackage[pdftex]{graphicx}
\usepackage{xcolor}

\begin{document}

\title{
A Distributed Ledger Based Infrastructure for Smart Transportation System and Social Good\footnotemark}

\author{\IEEEauthorblockN{Mirko Zichichi, Stefano Ferretti, Gabriele D'Angelo}
\IEEEauthorblockA{Department of Computer Science and Engineering\\
University of Bologna\\
Bologna, Italy\\
\{mirko.zichichi2, s.ferretti, g.dangelo\}@unibo.it}
}

\maketitle

\footnotetext{
The publisher version of this paper is available at \url{https://doi.org/10.1109/CCNC46108.2020.9045640}.
\textbf{{\color{red} This is the pre-peer reviewed version of the article: ``Mirko Zichichi, Stefano Ferretti, Gabriele D'Angelo. A Distributed Ledger Based Infrastructure for Smart Transportation System and Social Good. Proceedings of the IEEE Consumer Communications and Networking Conference 2020 (CCNC 2020).''.}}}

\begin{abstract}
This paper presents a system architecture to promote the development of smart transportation systems. Thanks to the use of distributed ledgers and related technologies, it is possible to create, store and share data generated by users through their sensors, while moving. In particular, IOTA and IPFS are used to store and certify data (and their related metadata) coming from sensors or by the users themselves. Ethereum is exploited as the smart contract platform that coordinates the data sharing and provisioning. The necessary privacy guarantees are provided by the usage of Zero Knowledge Proof. We show some results obtained from some use case scenarios that demonstrate how such technologies can be integrated to build novel smart services and to promote social good in user mobility.
\end{abstract}

\IEEEpeerreviewmaketitle

\section{Introduction}
In the last decades, smart transportation systems have emerged as a way to improve transportation efficiency, travel safety, vehicle security and better choices for drivers and passengers. 
Today, intelligent vehicles and transportation systems represent fundamental technologies, that improve drivers comfort and security. A variety of applications and protocols can be enforced altogether to obtain advanced and improved transportation systems. However, to fully exploit their potential and promote the development of smart mobility applications and services for social good, several novel challenges must be faced, that require substantial changes in transportation system models. The mentioned goals can be accomplished only through the use of procedures, systems and devices that allow data gathering, communication, analysis and distribution among individuals vehicles, infrastructures and services.

A reduced presence of (human) intermediaries can lead to the creation of smart services that take advantage of faster processing and better performances to provide the basis for smart moving and peer-to-peer services. Moreover, in the case of data sharing, service automation enables users to completely maintain control over the data they produce, making possible an intelligent sharing for social good.

In this scenario, another prominent technology that can play a main role is the decentralized management of crowd-sourced data, i.e.~the blockchain~\cite{D'Angelo201893}. The blockchain, made famous by Bitcoin~\cite{nakamoto2009bitcoin}, enabled a new vision for both finance and trust in distributed systems. Since their inception, the growth of blockchain technologies renewed the concepts of contracts and digital democracy, especially after the introduction of Ethereum~\cite{buterin2013ethereum}. The decentralized computation enabled by the Ethereum blockchain allows to create self-managed structures that do not rely on a central control, thus eliminating the presence of single point of failures~\cite{sf-gda}. Moreover, this blockchain allows using smart contracts in order to build Decentralized Applications (dApps) and Decentralized Autonomous Organizations (DAOs) that can realize novel important applications for social good~\cite{zichichi2019likestarter}.

Under the technical viewpoint, a blockchain is a specific type of Distributed Ledger Technology (DLT) with the scope to move trust from a human intermediary, that manages a transaction between two parties, to a protocol that allows the two parties to transact directly, i.e.~without the need of such third party. There are different implementations of DLTs, each one with its pros and cons. For example, Ethereum~\cite{buterin2013ethereum} provides a distributed virtual machine that is able to process any kind of computation but with constraints in scalability. Conversely, IOTA ledger~\cite{popov2016tangle} is thought to provide better scalability but it does not support distributed computation. Thus, if one wants to build a sophisticated software architecture, acting as the middleware for secure and certified smart transportation system applications, multiple DLTs can be utilized and combined, so as to take the best of multiple worlds. This is the philosophy we followed in our approach.

The aim of this work is to use DLTs to propose an infrastructure for smart transportation systems. Two main features are at the basis on this infrastructure: data sharing and smart services. We claim that the combination of data sharing and smart services allows creating a framework that promotes social good in user mobility. In particular, data sharing services are defined to let users and sensors to share their data. These services allow defining how the data can be shared but also how (from who and through which technology) they are acquired. The proposed infrastructure is based on DLTs, in combination with other technologies for distributed data management, i.e.~IPFS~\cite{benet2014ipfs}. Moreover, smart transportation and mobility services are defined and implemented through smart contracts.

In this work, we consider Vehicular Ad-hoc NETworks (VANETs) as the reference use case for smart transportation systems. In fact, VANETs allow vehicles to share information and to create a peer-to-peer substrate for the creation of social good applications in smart mobility environments. VANETs are classified as a specific type of Mobile Ad-hoc NETworks (MANETs) with the potential of improving road safety and travelers comfort~\cite{al2014comprehensive}. VANETs exploit wireless communication between moving vehicles, forming a landscape where vehicles can communicate between each other (V2V), with some fixed on-road equipment (V2X) or with the transport infrastructure (V2I).

In our architecture, vehicles generate data (e.g.~obtained through sensors) that can be distributed, stored and shared thanks to the combined use of IOTA and IPFS. Ethereum smart contracts provide the framework for the coordination among entities. Thus, smart contracts allow gaining access to specific data of interest, once the user has the authorization to access the data (e.g.~by paying for such access authorization). Zero Knowledge Proof is employed as the scheme to offer privacy while providing proof-of-location guarantees.

We present some results from an experimental campaign that demonstrate the viability of the proposed approach. A preliminary implementation of the key parts of the proposed architecture is freely available at \cite{git-app}.

The remainder of this paper is organized as follows. Section~\ref{sec:back} introduces some background and related work in the field. Section~\ref{sec:archi} presents the proposed distributed software architecture. Section~\ref{sec:eval} shows results from an experimental assessment. Finally, Section~\ref{sec:conc} provides some concluding remarks.

\section{Background and Related Work}\label{sec:back}

Our software architecture is essentially based on three main concepts: smart transportation systems, blockchain (or distributed ledgers) and Internet of Things (IoT). In this section, we quickly review some background aspects that are needed, as well as some related work in the field.

\subsection{Background}
In our proposal, data sharing is based on two decentralized technologies: IOTA and IPFS.

\subsubsection{IOTA}
IOTA is a permissionless DLT that allows hosts in a network to transfer immutable data and value among each other. It is specifically designed for the IoT industry. The ledger used in IOTA is not structured as a blockchain but as a Direct Acyclical Graph (DAG) called the Tangle~\cite{popov2016tangle}. In the IOTA DAG, the graph vertices represent transactions and edges represent approvals. When a new transaction is issued, it must approve two previous transactions and the result is represented by means of directed edges.

An important feature offered by IOTA is the Masked Authenticated Messaging (MAM). MAM is a second layer data communication protocol which adds functionality to emit and access an encrypted data stream over the Tangle. Data streams assume the form of channels, formed by the linked list of transactions in chronological order.  In other words, MAM enables users to subscribe and follow a stream of data, generated by some device. Our architecture makes an extensive use of this feature.

\subsubsection{IPFS}
The InterPlanetary File System (IPFS)~\cite{benet2014ipfs} is a protocol that allows to connect a set of computing devices in a peer-to-peer network sharing the same distributed file system. IPFS creates a resilient system of file storage and sharing with no single point of failure and in which the nodes do not need to trust each other. This technology is useful to store data that is not convenient to put on the DLTs.

\subsubsection{Zero Knowledge Proof}
To improve the privacy on the ledgers, we use a Zero Knowledge Proof~\cite{feige1988zero} protocol in which one party (called Prover) can prove to another party (called Verifier) that he knows a value $x$ without giving any information except the fact that he knows $x$ (without disclosing the $x$ value).

\subsubsection{Proof of Location}
Proof of Location (PoL) has been introduced recently to provide a location layer for smart contracts~\cite{migliorinienhancing} and it is used to state the correctness of a user's claim to be in a certain position at a given time. In this work, we use FOAM~\cite{foam2018}, an open protocol for decentralized and geospatial data markets in which PoLs are created using trustless devices. This mechanism is used in conjunction with Zero Knowledge PoL (zk-PoL)~\cite{wolberger2018zero}, that allows a Verifier to test whether position committed by a Prover is inside or outside the radius of a service area without revealing Prover's exact location.

\subsection{Related Work}
The blockchain has been recently widely adopted in the context of Internet of Things~\cite{wang2019survey}, smart cities~\cite{biswas2016securing,ibba2017citysense} and  Intelligent Transportation Systems (ITS). In~\cite{yuan2016towards}, the authors conduct a preliminary study of blockchain based ITS, giving the basis for a new ITS-oriented blockchain model. In this work, the focus is on the blockchain potential to help establish a secured, trusted and decentralized ecosystem, hence the advantages and research issues are showed to be used as reference in other works. In \cite{leiding2016}, the authors present to CHORUS Mobility, a a decentralized system that combines VANETs and Ethereum to provide services and enforce rules in smart transportation systems~\cite{leiding2018enabling}. In \cite{sharma2017block}, various use cases are shown by authors to validate blockchain based applications for social good. In environments such as Transportation Systems where data sharing is fundamental, how the data are obtained is crucial and could require a change of paradigms and technologies that are employed. These works \cite{ortega2018trusted,khelifi2018reputation} have a main goal which is similar to that of this paper, i.e.~the proposal of a blockchain based architecture where data is the main focus. In the case of data sharing restricted by payment, a basis for a marketplace is needed, such as the systems presented in \cite{shafagh2017towards,ozyilmaz2018idmob} that are backed by blockchains for IoT infrastructures and smart cities.

\section{System Architecture}\label{sec:archi}
\begin{figure*}[ht]
    \centering
	\includegraphics[width=.9\textwidth]{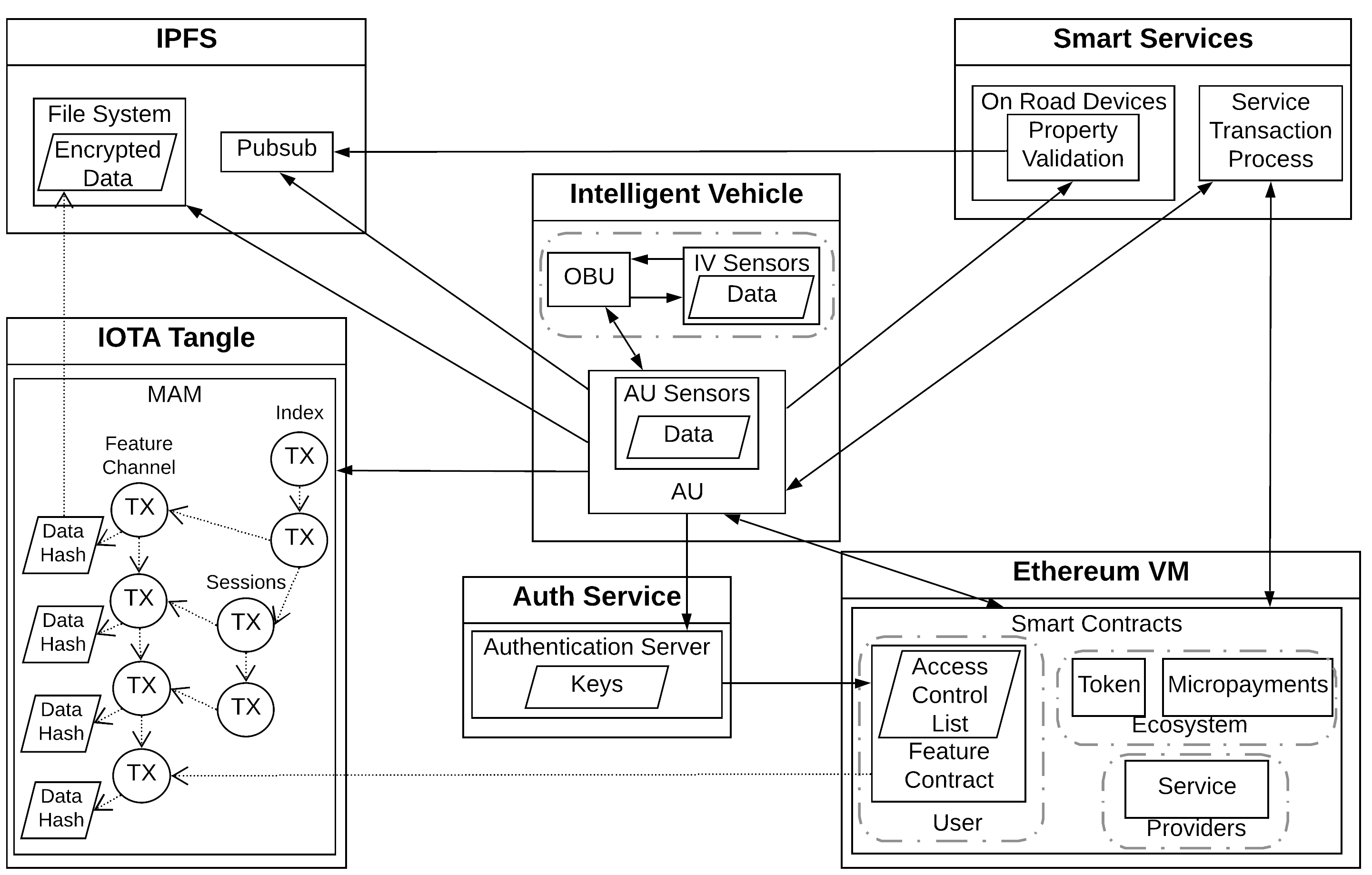}
	\caption{Infrastructure Architecture Diagram.}
	\label{fig:archdiagram}
\end{figure*}

Multiple distributed technologies are used in conjunction to provide data sharing and smart services capabilities. The peer-to-peer nature of Ethereum, IOTA and IPFS allows to combine them in a wider decentralized architecture and enables the nodes to communicate with other peers. The user operates through an Application Unit (AU), i.e.~his personal device, communicating with network peers and service providers. Through their AU, users share produced data and gain access to smart services (see Figure~\ref{fig:archdiagram}).

\subsection{Data Acquisition and Storage}
Data sharing is the feature at the core of the smart transportation system infrastructure. Currently, most data sharing services are based on central entities, that are in charge to control and authorize access to data. Clearly enough, these solutions inherit all the drawbacks of server-based approaches (e.g.~censorship, single point of failure).

In this work, data are managed and shared directly by the users using decentralized technologies to store and access them. In particular, different types of data can be acquired directly from the Intelligent Vehicle (IV) sensors or AU sensors, and then stored using the IOTA Tangle and IPFS.

\subsubsection{Data Production}
The user is the data producer. More specifically, data are obtained through the AU or the IV; in the latter case, sensed data can be transferred from the IV to the AU through an Intra-Vehicle communication technology (e.g.~Bluetooth, USB or Wi-Fi).

\subsubsection{IOTA Masked Authenticated Messaging}
The Tangle stores immutable information that cannot be censored/removed and will always remain online as long as the Tangle is online. Since everyone is able to access this information, it is necessary to obfuscate it to whom has no access rights. For this reason, in the proposed infrastructure MAM channels are used to store encrypted data, providing access only to eligible users. There are three different types of channels for each user:
\begin{itemize}
    \item \textbf{Feature Channels} - Data gathered from sensors are organized in features and each feature has its own MAM channel. A feature channel is a list of data or metadata of the same kind, arranged in a chronological order. It that can be considered as a log, in which each transaction has the same structure and contains the data itself or a reference (i.e.~hash pointer) to the data.
    \item \textbf{Session Channels} - Session Channels are updated each time a user starts a new session. Transactions in these channels store references to other feature channel transactions, created in that session.
    \item \textbf{Index Channel} - 
    The Index Channel is used to maintain the hierarchical structure, needed to reference all the user's channels.
    Transactions in this channel contain references to other channels and can be seen as equivalent to tree objects in a git repository: when a new channel is created then a new transaction is inserted in the Index Channel, containing the new reference.
\end{itemize}

\subsubsection{IPFS Objects}
If there is a small amount of data to be saved (e.g.~latitude and longitude coordinates) then the data are directly stored into a transaction (in the associated feature channel). Otherwise, the transaction contains a reference to an IPFS object (i.e.~the hash pointer of the IPFS object). Once an object is stored in IPFS, it becomes immediately available to all peers in the network, but an associated hash pointer must be inserted in the ledger, in order to certify the integrity of the data. Thus, the data is published as an IPFS object and then (asynchronously) referenced through its hash into a MAM transaction.

\subsection{Data Access Control}
Once data is stored in the Tangle or IPFS, everyone is able to access it achieving an agreement with the owner. These agreements are encoded using smart contracts that are decentralized intermediaries. The access to a specific datum or a feature channel, indeed, is purchasable using dedicated smart contract methods that enact monetary transactions in Ethereum. Thus, no direct interactions are needed among the data owner and users interested in his data. The keys needed for data access are managed (and released) by an Authentication Service.

\subsubsection{Features Contracts}
Feature contracts are smart contracts managed by users that regulate the access to data in feature channels. These contracts provide a list of the available types of access and the associated cost. The main structure consists of an Access Control List (ACL) that associates an Ethereum address to a bundle of data. This bundle is composed by addresses of IOTA transactions that contain a single datum (or a reference to an IPFS object) and by references to entire session or feature channels. In essence, the ACL represents the rights to access the user's data. Furthermore, these contracts are instantiated during the user ``registration'' thanks to the Factory design pattern, hence presenting always the same behavior.

\subsubsection{Authentication Service}
The Authentication Service is in charge of enforcing the access rights that are specified in the feature contract ACL and to release the access keys to who possess these rights. The architecture behind this process is the only one currently implemented as a client/server communication architecture. A server receives requests from different clients that want to access a specific user's data and answers providing a set of keys used to encrypt MAM Channels and IPFS Objects. Clearly, this happens only when the client is eligible for that data. The use of this service is necessary to unload the user's AU from the great amount of access requests, and also because it is unfeasible to encrypt a message through smart contracts without revealing the content of the message itself. (However, this is a limitation common to all DLTs, in general.) For each IOTA transaction and its possible related IPFS object, a key exists, which is used to encrypt the produced data. Such a key is obtained as the hash of the concatenation between a master key and the root of the MAM transaction: SHA256(master-key$\parallel$root). A transaction root is the root of a MerkleTree used in MAM transactions. Thus, a user only needs to communicate the master key and the Index Channel to the Service.

\subsection{Second Layer Trust}
DLTs are based on the trust in the consensus algorithm that governs the mechanisms underlying transactions validation, but correctness of data published on ledgers cannot be verified by these terms. Hence, it is necessary a second layer trust to accomplish this task. Through a Public Key Infrastructure (PKI), it is possible to verify the data correctness for intrinsic properties of the smart transportation system (e.g.~geospatial coordinates). More in general, Proofs or Certificates can be used to prove a certain user state property, and they can be attached together with data in feature channels.

\subsubsection{Proof of Location (PoL)}
One of the most important features needed in smart transportation systems is related to the geo-localization of sensors and data producers. In fact, knowing the user's spatial coordinates enables the implementation of context aware services. User's PoLs are produced by trusted devices or in a trustless way and they are used to verify data correctness. PKI permits any trusted device operating on road (e.g.~road side units or public transport vehicles) to release a PoL. Since communicating using WiFi (or Bluetooth) requires spacial closeness, when a user communicates through these technologies he can issue a claim to a ORD, that answers by releasing a signed certificate containing the device location. This certificate proves that the user is in the WiFi range of that ORD. In a trustless environment, such as that we are dealing with, we use FOAM to produce these PoLs, whereas the zk-PoL is used to maintain users' privacy, verifying users presence in a certain zone, without revealing their exact position.

\subsubsection{Publish/Subscribe Service}
An IPFS based pub/sub service is intended to provide a communication environment where users can share their certificates but also where they can be informed on traffic events and safety concerns. Using Zero Knowledge Proofs, users can produce certificates where they prove publicly in pub/sub channels what properties they own. A hierarchical zone organization of pub/sub channels can be dedicated to publish certificates, where users are able to detect other users to contact when they are interested to gather their data. This approach opens the way towards the implementation of a Data Marketplace for smart mobility and social good. 
However, the discussion of possible future work is out of the scope of this paper.

\subsection{Smart Services}
Smart Services have been developed by fully exploiting smart contract functionalities. The main focus here is on payments, which can happen both on-chain or off-chain. On-chain payments are standard transactions happening through Ethereum smart contract, where a unique token used for data sharing and smart services is moved between accounts. Off-chain payments consist of micropayments of the same token, happening in a direct communication channel between the two interested parties. These are regulated through a smart contract that manages the so called ``state channels'', in an implementation similar to $\mu$Raiden. Two kinds of smart services exist in the infrastructure:

\subsubsection{On Road Services}
On road services are direct communication services operated by devices/vehicles present in the smart transportation system. These use V2X connections, where X indicate a device that is capable of offering Smart Services and the connection happens through short range communications. 

Micropayments are used to pay these services. Examples of on road services are:
\begin{itemize}
    \item Parking and Bus tickets - Messages are exchanged through NFC to make a micropayment.
    \item Limited Traffic Area and Motorway tolls payments - Messages are exchanged through WiFi Direct passing through gates.
    \item Pay-per-Drive - Rented cars payed during the use, through the communication between the AU and the vehicle on board unit considering kilometers traveled.
\end{itemize}

\subsubsection{Smart Contract Based Services}
Users that are able to provide data or services can transact with other users through ad-hoc smart contracts. The service payment is accomplished directly on-chain, as soon as contract methods are called. Service providers can use data provided by users to acquire knowledge of a particular area. Such data foster the development of geo-localized smart services, built using smart contracts as business logic. Thus, for example, a user \emph{A} may be both a data provider and a service consumer, whereas a user \emph{B} provides a service by consuming data (that can be generated by users like \emph{A}). User \emph{B} gains access rights to some data through feature contracts owned by user like \emph{A} and then offers a service based on that data to \emph{A}. Such service might be offered in exchange of a payment; the payment will be accomplished through a transaction issued to a specific smart contract related to the service and owned by \emph{B}.

\section{Performance Evaluation}\label{sec:eval}
Data sharing covers almost the entire infrastructure, hence performance evaluation must assess whether these sharing mechanisms present bottlenecks. Assuming that it is feasible to deploy enough IPFS nodes in order to let anyone store his data, the operation of data storing in IOTA may represent a bottleneck. Publishing transactions (TXs) in MAM channels is, indeed, the primary action that leads to all the benefits offered by this smart infrastructure. 

In this section, latency measurements are evaluated in different contexts and compared with the aim to provide a prospective on actual IOTA network capabilities. The results will be showed in plots and described together with the cases they bring up.
The process of attaching a TX to the Tangle includes two sub-processes:
\begin{itemize}
\item \textbf{Tips selection} - To attach a TX to the Tangle, it is mandatory to reference two other TXs called tips. These tips are provided by the IOTA full node that stores the Tangle.
\item \textbf{Proof-of-Work (PoW)} - This step consists in performing the PoW for the TX in order to be validated. 
In out tests, this task is performed by the IOTA full node.
\end{itemize}

To perform the evaluation assessment, a personal computer (PC) and a smartphone (AU) with mobile connection (4G) have been used. 
IOTA nodes may operate in two kinds of networks which differ for PoW difficulty (i.e.~the number of zeros at the end of a TX hash). Two providers are tested on the Mainnet with PoW difficulty of 14 (\textit{http://node.deviceproof.org} referenced as Provider1 and \textit{https://nodes.thetangle.org} as Provider2) and only one in the Devnet with PoW difficulty of 9 (\textit{https://nodes.devnet.iota.org}).
\subsubsection{IOTA full nodes performances}

\begin{figure}[th]
    \centering
	\includegraphics[width=.9\textwidth]{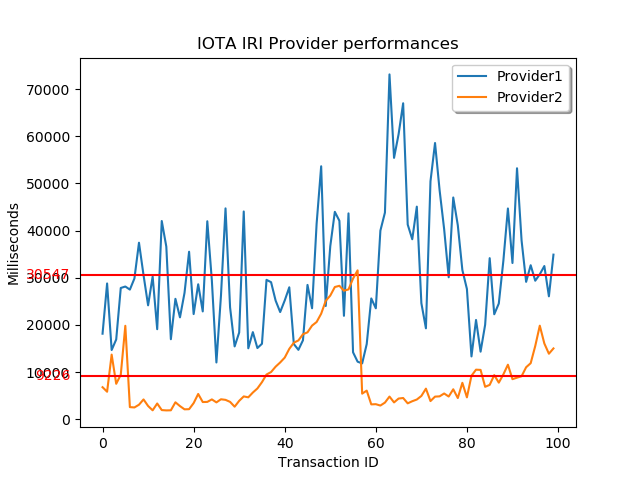}
	\caption{Latency comparison between two IOTA providers.}
	\label{fig:val-nodes}
\end{figure}

An important factor to consider in this evaluation is the performance gap between IOTA full nodes. As shown in Figure~\ref{fig:val-nodes}, the same task may be performed differently by two different providers. This is usually caused by the difference in resources and to the fact that a node with a higher network degree may receive more TXs from others; therefore, it is faster to provide tips to clients. This test consists in attaching 100 TXs to the Tangle Mainnet, but \emph{Provider1} is relatively ``unknown'', while \emph{Provider2} is one of the most used node that operates in IOTA. This means that \emph{Provider1} has more resources to provide to clients but less tips, since it receives less TXs, while \emph{Provider2} provides tips faster but has a limited amount of resources for each client. The result is that \emph{Provider2} allows to attach TX in 9 seconds on average, while \emph{Provider1} needs more that 30 seconds. The downside of \emph{Provider2} is that it allows a limited number of requests per time fraction, due to its limited resources. Indeed, during the tests \emph{Provider2} has shown a lower level of availability. This is maybe due to some specific policies of \emph{Provider2}. In particular, in our tests we encountered problems after $\sim 30$ MAM transaction messages, on average. (Note that a MAM transaction message corresponds to a bundle of 4 normal transactions). After an initial amount of successful MAM transactions, the network discarded all the subsequent ones. Probably, the net inserts particularly active nodes in a blacklist, in order to share resources among all the nodes. In our tests, the issuer node sent 0.4 transaction/sec. This rate is evidently too high. The presence of such blacklist might be exploited to avoid Denial of Service attacks as well. In any case, this problem deserves some further investigation.

\subsubsection{Comparison between PC and AU}
\begin{figure*}[ht]
    \centering
	\includegraphics[width=.9\textwidth]{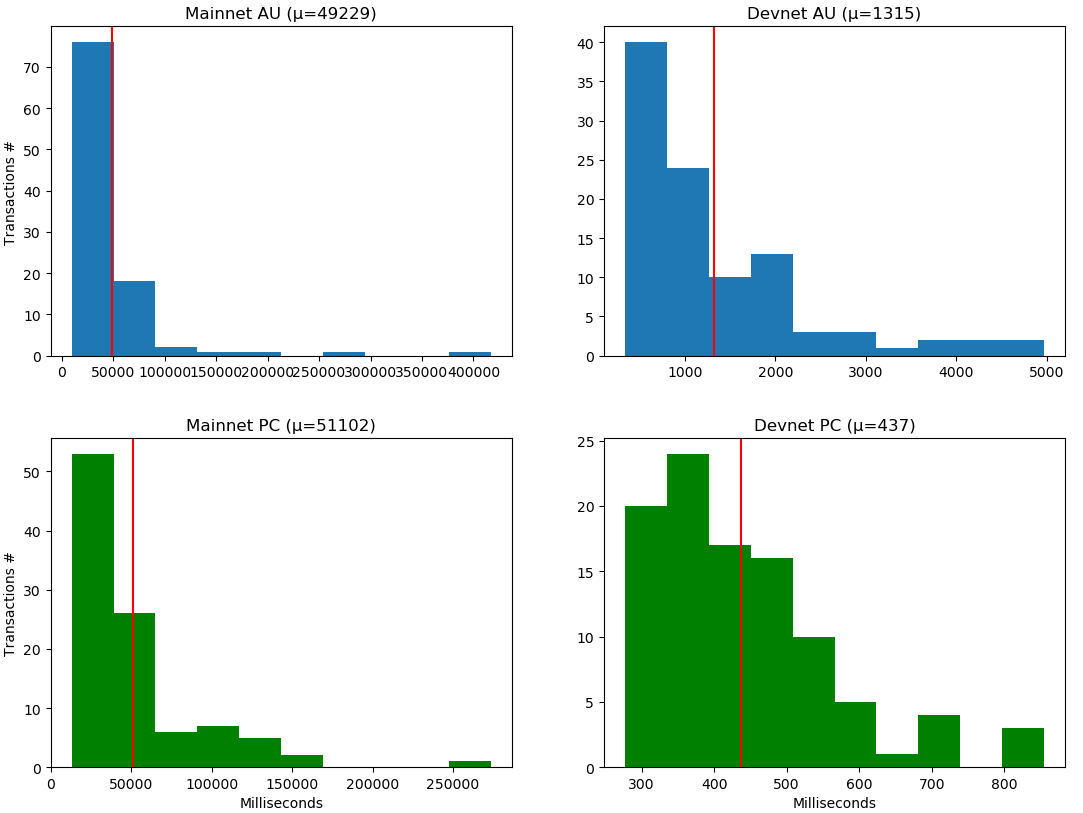}
	\caption{Histograms of latency in the process of MAM TX attachment to the tangle for PC and AU in Mainnet and Devnet.}
	\label{fig:val-compa}
\end{figure*}
The main evaluation consists in the comparison of latencies to attach a MAM TX, when varying the device exploited to transmit the transaction (i.e.~PC or AU). This experiment allows to assess the viability on using IOTA through mobile devices. 
In particular, this test consists in attaching 100 TXs to the same MAM channel, where each TX results in a bundle, containing the overhead structure needed for the MAM. The creation of such structure requires, on average, 1475 msec with the AU, while 224 msec with the PC.
In Figure~\ref{fig:val-compa}, histograms on the left show the latency when working in the Mainnet, while those on the right show latencies in the Devnet. In the case of the Mainnet, there is almost no difference between latencies with the AU and PC, that are on average around 50 sec. On the other hand, due to the decreased PoW difficulty in the Devnet, it takes, on average, only 437 msec using the PC and 1315 msec using the AU. The difference here may be attributed to load bursts. Devnet provider is, indeed, one of the most used, such as \emph{Provider2} in Mainnet, but its load is lower due to the minor amount of requests received, that is quite typical in a Devnet.

These results allow to conclude that performances on AU and PC are similar.

\section{Conclusion}\label{sec:conc}
In this paper, we presented a system architecture that exploits data generated in the novel generation of transportation systems, and allows building novel smart transportation services. To this aim, different DLTs are exploited. IOTA and IPFS are used as the backbone to store and share sensed and produced data. Ethereum smart contracts are exploited to control data access and authorization. Through this approach, it is possible to share data and elaborate them, hence paving the way for novel smart services for social good.

We provided some results on tests on the IOTA blockchain, which is the main technology for the data sharing, to investigate if this DLT can represent a bottleneck in the system. Results seem to confirm the viability of the proposed approach.

\section*{Acknowledgment}
This project has received funding from the European Union’s Horizon 2020 research and innovation programme under the Marie Skłodowska-Curie ITN EJD grant agreement No 814177.

\bibliographystyle{IEEEtran}
\bibliography{IEEEabrv,References.bib}

\end{document}